# Frequency-Multiplexed bias and readout of a 16-pixel Superconducting Nanowire Single-Photon Detector Array.

S. Doerner,[1,a)] A. Kuzmin,[1] S.Wuensch,[1] I.Charaev,[1] Florian Boes,[2] Thomas Zwick[2] and M. Siegel[1]

[1]*Institut für Mikro- und Nanoelektronische Systeme, KIT, Hertzstr. 16, 76187, Karlsruhe, Germany*
[2]*Institut für Hochfrequenztechnik und Elektronik, KIT, Engesserstr. 5, 76131, Karlsruhe, Germany*

**Abstract:** We demonstrate a 16-pixel array of radio-frequency superconducting nanowire single-photon detectors with an integrated and scalable frequency-division multiplexing architecture, reducing the required bias and readout lines to a single microwave feed line. The electrical behavior of the photon-sensitive nanowires, embedded in a resonant circuit, as well as the optical performance and timing jitter of the single detectors is discussed. Besides the single pixel measurements we also demonstrate the operation of a 16-pixel array with a temporal, spatial and photon-number resolution.

Superconducting Nanowire Single-Photon Detectors (SNSPDs) play an increasing role in applications like light detection and ranging (LIDAR)[1], quantum optics[2,3], long-distance optical communication[4], or time-resolved spectroscopy[5]. Many of these applications require multi-pixel devices in order to reduce analysis time or to achieve a photon-number resolution. For that reason, SNSPD arrays have been developed during the last years. These arrays require a multiplexing scheme to reduce the increasing system complexity and heat load due to high-pixel numbers. The state-of-the-art multiplexing schemes are based on a rapid single-flux quantum logic[6], time-tagged readout[7,8] or current splitting techniques[9]. However, none of those approaches is able to multiplex the bias lines without losing the adjustability of the bias current for each pixel, individually. In addition to SNSPDs, superconducting transition-edge sensors (TES) as well as microwave kinetic-inductance detectors (MKIDs) also provide a single-photon resolution in a wide spectral bandwidth[10,11]. Furthermore, they can be operated in large arrays using frequency-division multiplexing (FDM) readout schemes[12,13]. However, their response in the single-photon regime is rather slow and merely in the range of μs[10,11]. To combine the advantages of classical SNSPDs, especially the ps-range temporal resolution[14], with the intrinsic FDM capability of MKIDs we developed the Radio-Frequency Superconducting Nanowire Single-Photon Detector (RF-SNSPD)[15].

The RF-SNSPD consist of the classical meander-shaped SNSPD which is embedded into a lumped-element resonator. The current, oscillating in this resonant circuit at certain resonant frequency, is used to bias and readout the detector. Thus, the detector can be easily multiplexed in the frequency domain like MKIDs. However, the operation of the RF-SNSPD differs drastically from the MKID. To enable single-photon detection the RF-SNSPD is driven by a strong microwave signal with amplitudes close to the critical current, $I_C$, of the meander, which has a significantly smaller current-carrying cross-section. In this highly nonlinear regime the absorption of a single-photon is sufficient to create a normal conducting domain inside the meander which increases the damping of the resonator and rapidly stops any oscillation. Since it could be also considered as a reduction of the quality factor of the RF-SNSPD, the timing resolution is in the same order of magnitude as conventional SNSPDs. In contrast to MKIDs, where a slight change in the kinetic inductance due to photon absorption should be carefully amplified and detected, the switching of the resistance due to photon absorption in the RF-SNSPD is much stronger than the readout noise and allows operation at 4.2 K.

In this paper, we present results, of the development of a 16-pixel RF-SNSPD array. All detectors are coupled to one common feed line. This reduces the number of wires between the readout electronics at room temperature and the

a)Electronic mail: steffen.doerner@kit.edu

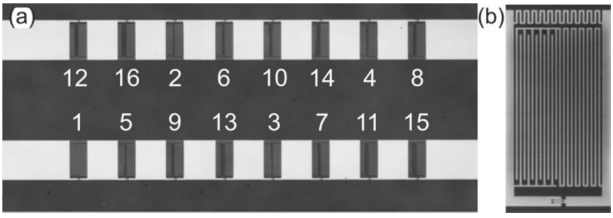

FIG 1 (a) Optical microscope image of a 16-pixel array. The corresponding pixel numbers are labeled on the inner conductor of the CPW. (b) SEM image of pixel number 1.

cryogenic stage to a minimum of two coaxial cables. The FDM method enables readout of all 16 pixels at the same time with a time, spatial and photon-number resolution. Furthermore, the FDM allows adjustable individual bias levels for each detector. To the best of our knowledge there have been no experiments, which are demonstrating bias-line multiplexing or the operation of a SNSPD array with time, spatial and photon-number resolution requiring less wiring.

In figure 1(a) a 16-pixel array is shown. All RF-SNSPDs are placed in the gaps of one common coplanar waveguide (CPW) and are coupled in parallel to the inner conductor and the ground planes. On figure 1(b) an SEM image illustrates the layout of a single pixel. The resonant frequency $f_{res}$ is mainly defined by a parallel lumped element LC circuit, consisting of a meandered inductor $L_p$ and interdigital capacitor $C_p$ with a line width $w > 1$ μm. This resonator is coupled to the CPW using an interdigital capacitor on the one side and the photon sensitive meander with $w < 100$ nm on the opposite. The inductance of $L_P$ is varying among the pixels starting from 55 pH for pixel 1 and decreasing down to 24 pH in pixel 16. The capacitance of $C_P$ is fixed for all devices at 54 fF. A more detailed explanation of the used resonant circuit can be found in our previous article.[16]

To obtain a spatial resolution of the array, we matched each RF-SNSPD to a certain $f_{res}$. Therefore, we tuned the inductance of the parallel LC circuit of each pixel to get a frequency spacing of 85 MHz between two detectors which results in a total bandwidth of 1250 MHz to operate all pixels (FIG 2). All other parts of the RF-SNSPDs, especially the photon sensitive nanowires, are the same in each structure. In figure 2 the simulated transmission parameters over frequency of the CPW are shown. For the simulation we used *Sonnet em* software[17] and took the properties of 330 μm thick R-plane sapphire ($\varepsilon_r$ = 10.06 and $\tan\delta$ = $10^{-9}$) as substrate into account. As superconducting material 4 nm thick NbN with a surface inductance of 55 pH/□ is used. To prevent crosstalk, we arranged the pixels among the array, so that two pixels facing each other on the chip are not adjacent in frequency. Furthermore, two neighboring pixels in frequency are alternately placed in the upper and lower gap of the CPW. The position of each pixel in the frequency range and on the chip is given in figure 1 and 2.

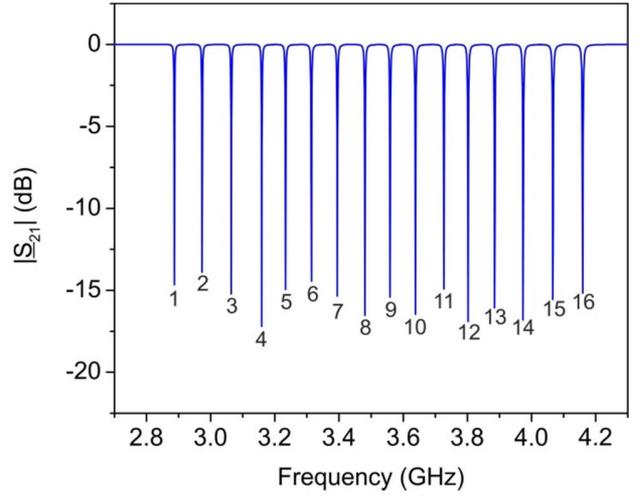

FIG 2 Simulation of the transmission over frequency of the 16-pixel array. The pixel numbers, corresponding to the transmission dips, are labeled below.

The RF-SNSPD requires only one single NbN layer which is used for the resonant circuit as well as for the photon sensitive nanowire. Thus, it is easy to fabricate the RF-SNSPD in the standard SNSPD fabrication process. At the first step, the 4-nm-thick NbN is deposited on one-side polished R-plane sapphire substrates at a temperature of 850°C using reactive magnetron sputtering. Second fabrication step is the patterning of the CPW and the RF-SNSPDs using electron-beam lithography with PMMA 950k resist. The final step is the transfer of the patterned geometry into the NbN layer using Ar-ion beam etching.

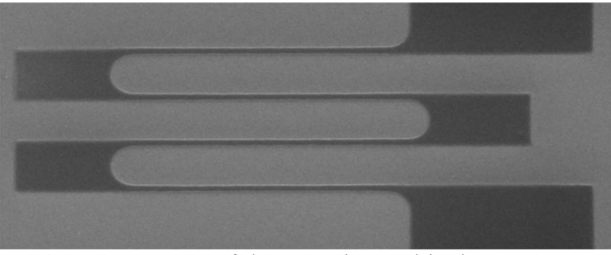

FIG 3  SEM Image of the nanowire used in the RF-SNSPDs. The overall length of the wire is 17μm and the width is 73 nm.

Figure 3 shows one of the fabricated nanowires with a length of 17 μm and $w$ = 73 nm. We designed the radius of the nanowire's bends to be rather big to avoid any influence of the current crowding effect[18], which later results in a filling factor of 14%. In future devices the filling factor could be increased using spiral nanowires[19].

We measured the microwave transmission spectrum of the fabricated array, which is shown in figure 4. The chip was mounted in a gold-coated brass housing, equipped with 2 coaxial adapters to feed through microwave signals. The housing also allows optical illumination of the detector through a multimode fiber. The input microwave signal is attenuated prior the array by 32 dB using a cold attenuator to reduce external noise. The output signal is boosted by a broadband low-noise HEMT amplifier. The setup was mounted in a dipstick and cooled down to 4.2 K using a transport $^4$He dewar. The transmission spectrum of the sample was recorded in the relevant bandwidth using a vector network analyzer (VNA). Appling low power levels to the CPW in the bandwidth of the array, all resonances of the detectors are visible (FIG 4). At a certain level of the microwave power, $P_C$, the oscillating current inside each resonator reaches the critical current of the nanowire and the meander switches to the normal-conducting state. The transmission spectrum for a level above $P_C$, which is depicted in figure 4, shows that all RF-SNSPDs are functional. The strongly increased damping of the resonator in this case stops the oscillation.

In order to tune each resonance frequency, we change the length of the inductor $L_P$, assuming a constant surface inductance of 55 pH per square. But, the current distribution in the bends of the inductor is non-uniform[18]. Thus, the non-equal frequency spacing between the pixels in figure 4 can be explained by the varying number of bends in the inductor $L_P$ among the pixels. Some pixels differ only in the length of $L_P$, where some others differ in the length and in the number of bends. To check if the designed relative position of each detector in the array corresponds to the designed position in the frequency domain, we measured the transmission spectrum sequentially shorting each RF-SNSPD with an indium bond. The result of the measurements matches the expected relative positions of the pixels. During all transmission measurements, we did not see any crosstalk between any pixels. The increased damping for higher frequencies in figure 4 results from the uncalibrated setup.

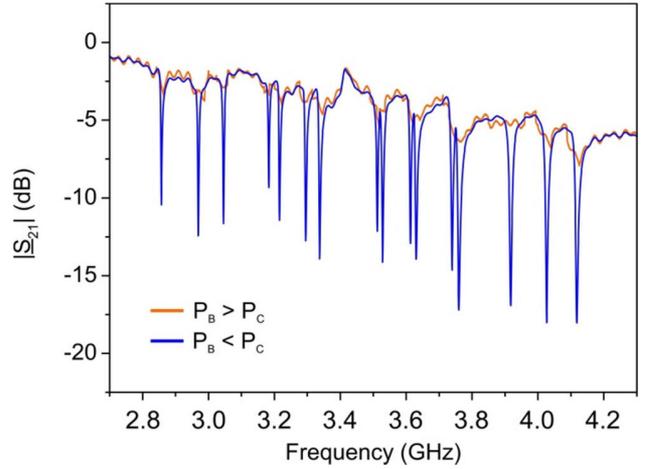

FIG 4  Measured transmission over frequency of the 16-pixel array. We performed one measurement well above and one well below the critical power level of each device.

For the optical measurements, the VNA was replaced with an analog signal generator, which produces the single tones to bias and probe the RF-SNSPDs. A 32-GHz real-time oscilloscope, which measures the time-resolved detector response, was used as back-end. This setup was used to record the change of the transmission due to a single-photon detection event of the RF-SNSPD. The example of such an event is depicted in figure 5 for detector number 16. At $t < 0$ the detector is in the superconducting state. At $t = 0$ an instantaneous change in the transmission can be observed, which is caused by the formation of a normal-conducting domain in the nanowire after photon absorption. The rise of the transmitted signal is about 11 dB, as one would expect from the depth of the resonance, measured on VNA (FIG. 4). This instantaneous change of the quality factor to almost zero allows very high timing precision of the detector. At t > 0 the nanowire recovered the superconducting state and the transmitted signal starts to decrease according to the quality factor of the RF-SNSPD which mainly limits the reset time after photon detection.

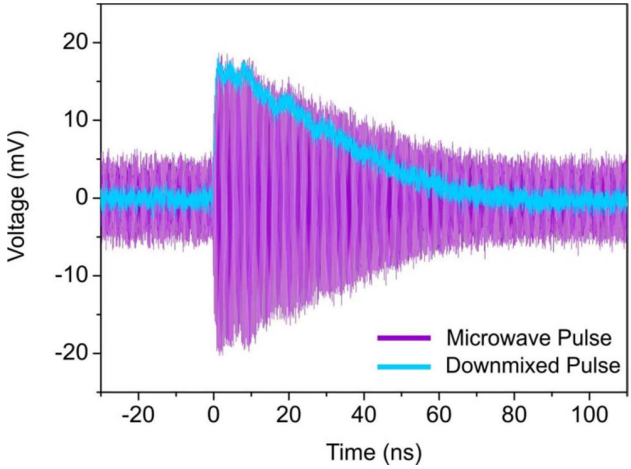

FIG 5 Measured voltage over time on the feed line. The microwave signal, measured after the first amplifyer as well as the down-mixed microwave signal using the IQ-mixing setup is shown. For a better comparison, the voltage levels of the down-mixed pulse is scaled to the measured microwave pulse.

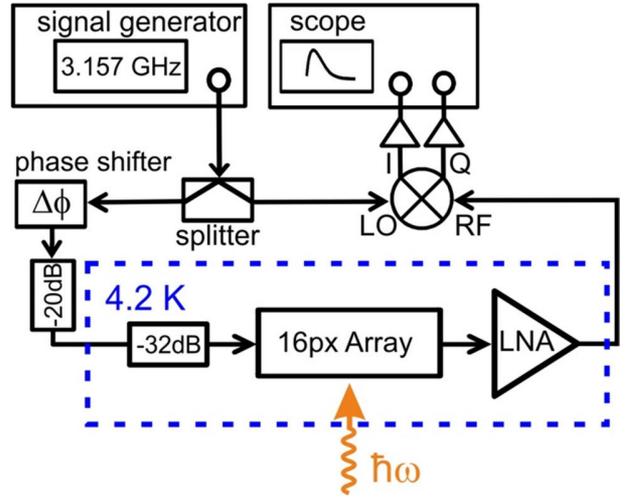

FIG 6 Schematic of our setup used to down mix the amplitude modulation of the carrier frequency after a detection event.

To quantify the timing resolution, we performed a jitter measurement using a pulsed femtosecond laser at 1.55-µm wavelength. We statistically analyzed the delay times between the detector responses and the correspondent reference pulses of the laser. As a detector response we used the amplitude change of the complex transmission. To perform fast measurement of the complex-transmission transient we used a phase-sensitive homodyne scheme similar to MKID readout schemes. It includes a signal generator and IQ mixer for the demodulation of the carrier signal. To eliminate the quadrature component Q at the output of the IQ mixer, the phase of the input microwave carrier was adjusted with a phase shifter. The 32-GHz oscilloscope is used as the back-end to record the I-component of the transient, which represents the count events of photons. A scheme of our setup is given in figure 6. Recording a delay time distribution of more than $10^4$ events, we measured a typical detector jitter of 59 ps at the full width at half maximum.

The optical performance of each detector was characterized individually, measuring their pulse-count rates. Each detector was biased with 80% of the critical-power level. The wavelength of the incident photons have been varied during each measurement from 400 nm up to 750 nm. The detection efficiency (DE) was afterwards calculated as the ratio of the measured count rate to the number of incident photons onto the nanowire. The measured results showed a single-photon sensitivity of each pixel with a maximum detection efficiency ranging from 2% up to 10%. We also tested the performance of the array operating multiple pixels simultaneously. The signal generator was replaced by an arbitrary waveform generator, which is able to produce eight tones with a specific power level at once. The readout of the array is similar to other FDM schemes and based on the observation of an amplitude change at the corresponding frequencies. A detailed explanation of the readout method is given in our previous article[16], where we demonstrated the crosstalk-free operation of a two-pixel RF-SNSPD array with photon-number resolution. Using this setup, we were able to record the spot profile of the fiber tip above the array in only two measurements. The results are shown in figure 7. The center of the spot was adjusted on the first pixels of both rows where the maximum count rate was measured. With increasing detector index the count rate drops down. It does not drop down to zero because the spot diameter is much larger than the array. Our setup does not allow the illumination of single pixels by keeping others dark. To get the proper beam distribution we normalized the count rates to the measured detection efficiencies.

In summary, we have successfully demonstrated the operation of a 16 pixel RF-SNSPD array with an integrated and scalable multiplexing scheme. The bias and readout signals are multiplexed in the frequency domain which allows the operation of each pixel at an individual bias level using only one common microwave feed line. We showed the photon sensitivity of each pixel as well as the time, spatial and photon-number resolution of the array. Furthermore, we demonstrated the timing accuracy of the detected single photons with a precision of 59 ps.

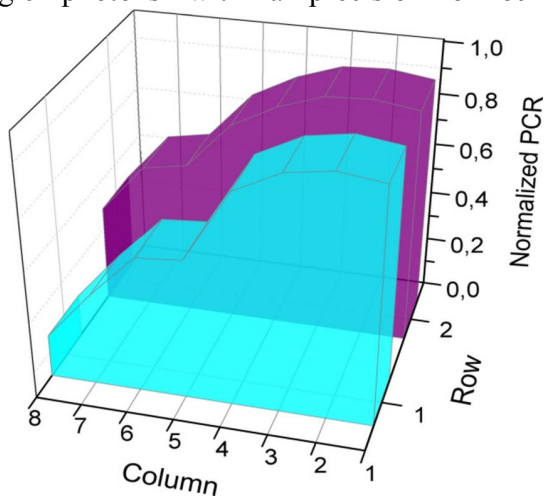

FIG 7  Normalized average pulse count rate of the 16-pixel array. On the right side the pixels are in the center of the spot, why the maximum count rates are measured. To the left side the count rates show the decaying of the spot intensity.

This work was supported in part by the Karlsruhe School of Optics and Photonics (KSOP).